\documentclass[twocolumn,showkeys,aps,prl,showpacs]{revtex4-1}
\UseRawInputEncoding
\usepackage{graphicx}
\usepackage[CJKbookmarks,dvipdfm,colorlinks,linkcolor=blue,citecolor=blue]{hyperref}
\usepackage{amsmath, amssymb}
\usepackage{makecell}
\usepackage{multirow}
\usepackage{CJK}
\usepackage{mathrsfs}
\usepackage{bm}
\usepackage{amsmath}
\usepackage{dcolumn}
\usepackage{epstopdf}
\usepackage{dsfont}
\usepackage{amssymb}
\usepackage{tabularx}
\usepackage{array}
\usepackage{float}
\usepackage{color}
\usepackage{epstopdf}
\usepackage{mathrsfs}
\usepackage{extarrows}

\begin{document}

\title{How to produce spin-splitting in antiferromagnetic   materials }

\author{San-Dong Guo}
\email{sandongyuwang@163.com}
\affiliation{School of Electronic Engineering, Xi'an University of Posts and Telecommunications, Xi'an 710121, China}
\author{ Guang-Zhao Wang}
\affiliation{School of Electronic Information Engineering, Yangtze Normal University, Chongqing 408100, China}

\author{Yee Sin Ang}
\email{yeesin_ang@sutd.edu.sg}
\affiliation{Science, Mathematics and Technology (SMT), Singapore University of Technology and Design, Singapore 487372}
\begin{abstract}
Antiferromagnetic (AFM) materials have potential advantages for spintronics due to their robustness, ultrafast
dynamics, and magnetotransport effects.  However, the missing spontaneous polarization and magnetization hinders the efficient utilization of electronic spin in these AFM materials. Here, we propose a simple way to produce spin-splitting in AFM  materials
by making the magnetic atoms with opposite spin polarization locating in the different environment (surrounding atomic arrangement), which does not necessarily require the presence of spin-orbital coupling (SOC).
We confirm our proposal by four different types of two-dimensional (2D) AFM materials within the first-principles calculations.
 Our works provide a intuitional  design principle to find or produce  spin-splitting in AFM   materials.

\end{abstract}
\keywords{Spin-splitting, Antiferromagnetism ~~~~~~~~~~Email:sandongyuwang@163.com}

\maketitle

\section{Introduction}
 The spin-orbit-coupling (SOC)-induced spin splitting have formed
the basis for the development of spintronics within potential applications to spin transistor, spin-orbit
torque, spin Hall effect, topological insulators, and Majorana
Fermions\cite{ss1}.  However, the rapid decoherence  of spin-polarized electrons induced by SOC limit the widespread applications
of these materials\cite{ss2}.The unidirectional spin-orbit field orientation can
produce a spatially periodic mode of the spin polarization,  which is  known as the persistent spin helix (PSH)\cite{p7,p8}. The PSH can give rise to an extremely long spin lifetime by suppressing  spin dephasing  due to SU(2) spin rotation symmetry\cite{p7,p9}.
To produce obvious spin splitting, these materials should  incorporate heavy elements with the rarity, instability and
toxicity. So,  an alternative strategy should be used  to realize spin splitting in the absence of SOC.

\begin{figure}
  \includegraphics[width=7cm]{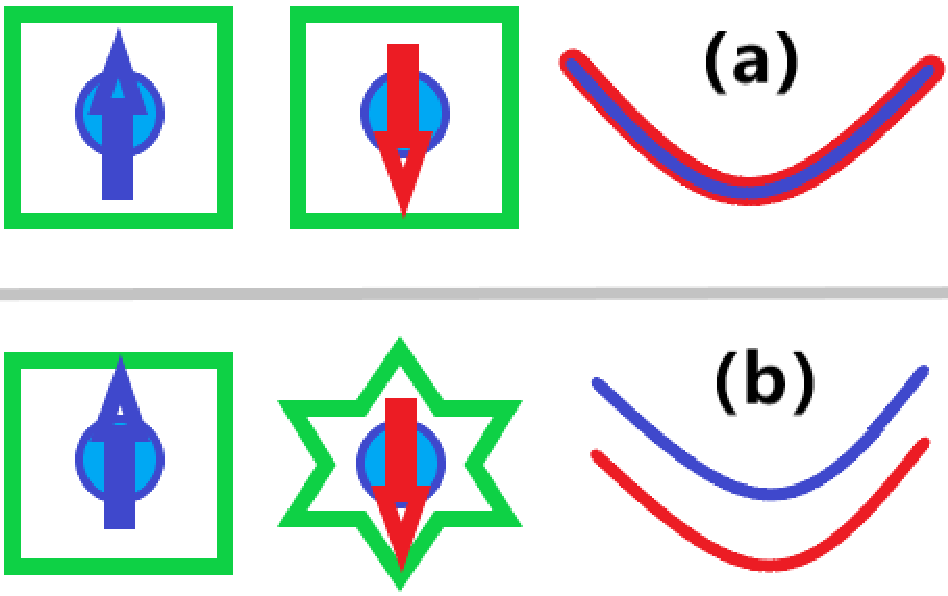}
  \caption{(Color online) For antiferromagnetic   materials,  (a):  the magnetic atoms with opposite spin polarization have the same environment (surrounding atomic arrangement), producing the degeneration of electron spin ; (b): the magnetic atoms with opposite spin polarization have different environment, destroying the degeneration of electron spin. }\label{sy}
\end{figure}

The Zeeman spin splitting in ferromagnets  is irrelevant to SOC,  and  can be produced even
when SOC is turned off. Any magnetic compound that has nonzero net magnetization can give rise to spin splitting with underlying Zeeman mechanism.
A typical example is half-metallic ferromagnet, which possesses   100 \% spin-polarized current\cite{ss3}. Superior to ferromagnetic
(FM) materials, the  antiferromagnetic (AFM)  materials  have attracted considerable research interest, since they  are robust to external
magnetic perturbation due to missing any net magnetic moment\cite{k1,k2}.
In general, there are not   spin splitting  in the band
structures  in these antiferromagnets.

\begin{figure*}
  \includegraphics[width=14cm]{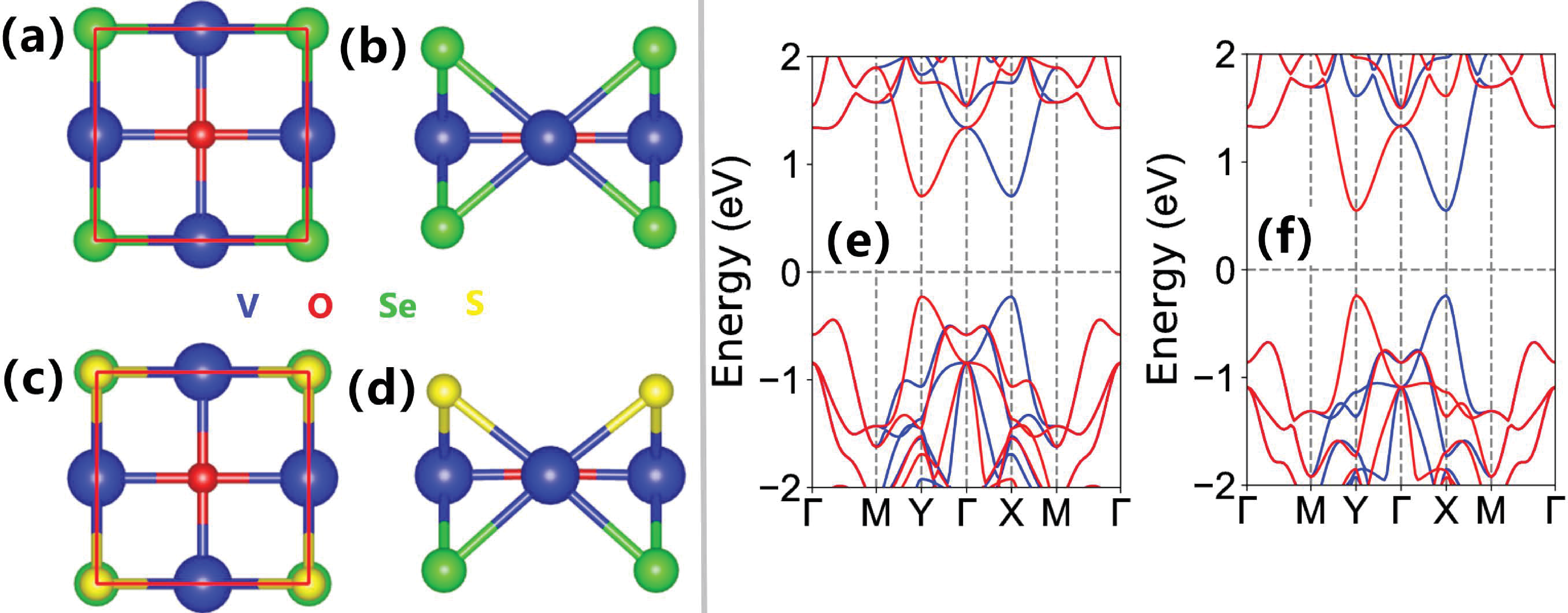}
  \caption{(Color online)  The top (a,c) and side (b,d) views of the  crystal structures for  $\mathrm{V_2Se_2O}$ (a,b) and $\mathrm{V_2SSeO}$ (c,d). The energy band structures of  $\mathrm{V_2Se_2O}$ (e) and $\mathrm{V_2SSeO}$ (f), and the spin-up
and spin-down channels are depicted in blue and red.}\label{st1}
\end{figure*}
\begin{figure}
  \includegraphics[width=5cm]{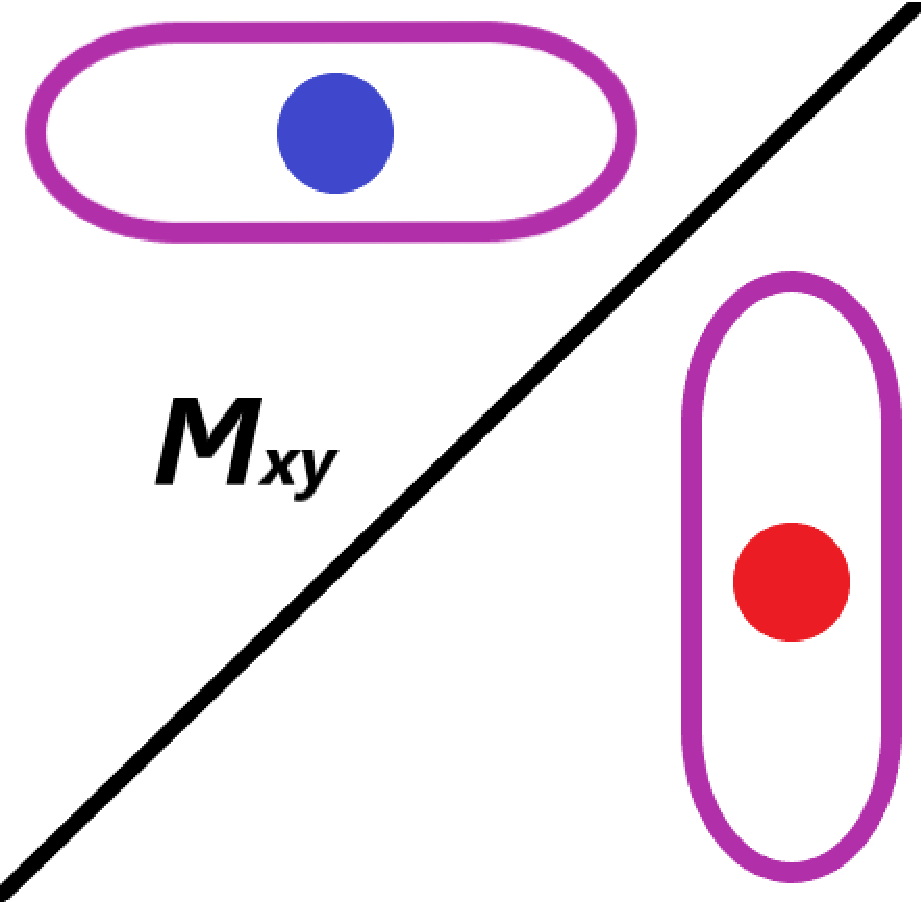}
  \caption{(Color online)The two V atoms have the same surrounding Se atomic arrangement as a rectangle, but  the directions  of the two rectangle are different ($x$ and $y$ directions), producing the spin splitting. Two  V-atom related  sublattices  is related by a diagonal mirror symmetry $M_{xy}$, which leads to that two spin channels in the band structures is also related by $M_{xy}$.  }\label{st1-1}
\end{figure}

However,  the altermagnetism  with collinear symmetry-compensated  antiferromagnetism can realize
the spin splitting, where  the SOC  is  not needed\cite{k4,k5,k6}. The altermagnetism is characterized by crystal-rotation
symmetries connecting opposite-spin sublattices separated in the real space, which leads to opposite-spin electronic
states separated in the momentum space.
Several bulk materials and two-dimensional (2D)  materials have been predicted to be  altermagnetism, such as $\mathrm{RuO_2}$\cite{k7}, $\mathrm{FeF_2}$\cite{k7-1}, MnTe\cite{k7-2},
some organic AFMs\cite{k8}, $\mathrm{MnF_2}$\cite{k9}, some $\mathrm{GdFeO_3}$-type
perovskites\cite{k10},  $\mathrm{Cr_2O_2}$\cite{k11,k12}, $\mathrm{Cr_2SO}$\cite{k12-1}  and  $\mathrm{V_2Se_2O}$\cite{k13}.
 Recently,  a new
mechanism has been proposed to achieve spin splitting in  AFM materials\cite{k14}. For a 2D material, the magnetic atoms have opposite layer spin polarization (A-type AFM ordering) with an out-of-plane built-in electric field, which can  destroy the degeneration of electron spin in the band structures, called electric-potential-difference antiferromagnetism (EPD-AFM). Janus monolayer $\mathrm{Mn_2ClF}$  is proved to be a possible candidate to achieve the EPD-AFM by the first-principles calculations\cite{k14}.

Is there an intuitive way to search for or produce spin splitting in AFM materials? Here,  a simple way is proposed to produce spin-splitting in AFM  materials. For antiferromagnetic   materials: if  the magnetic atoms with opposite spin polarization have the same environment (surrounding atomic arrangement), the degeneration of electron spin can be observed (see \autoref{sy} (a)); if the magnetic atoms with opposite spin polarization have different environment,  the degeneration of electron spin can be removed (see \autoref{sy} (b)).
For  altermagnetism, the two different environments can be related by special symmetry operation, which leads to that two spin channels in the band structures is also related by the corresponding symmetry operation, for example 2D $\mathrm{Cr_2O_2}$\cite{k11,k12}, $\mathrm{Cr_2SO}$\cite{k12-1}  and  $\mathrm{V_2Se_2O}$\cite{k13}. For EPD-AFM, the different  environments occupied by two magnetic atoms are induced by electric-potential-difference, for example Janus monolayer $\mathrm{Mn_2ClF}$\cite{k14}.

The rest of the paper is organized as follows. In the next
section, we shall give our computational details and methods.
In  the next few sections, four different types of 2D AFM materials ($\mathrm{V_2SSeO}$,  $\mathrm{Mn_2ClI}$,  $\mathrm{CrMoC_2S_6}$ and  $\mathrm{V_2F_7Cl}$) are used to confirm our idea. Here, we are not particularly concerned with the possibility of experimental synthesis of these materials. Finally, we shall give our discussion and conclusions.

\section{Computational detail}
 The spin-polarized  first-principles calculations are performed  within density functional theory (DFT)\cite{1} by using the projector augmented-wave (PAW) method,  as implemented in VASP code\cite{pv1,pv2,pv3}. The generalized gradient
approximation  of Perdew-Burke-Ernzerhof (PBE-GGA)\cite{pbe} is adopted  as the exchange-correlation functional. To account for electron correlation of $d$ orbitals,  a Hubbard correction $U_{eff}$ is adopted  within the
rotationally invariant approach proposed by Dudarev et al.
The kinetic energy cutoff  of 500 eV,  total energy  convergence criterion of  $10^{-8}$ eV, and  force convergence criterion of less than 0.001 $\mathrm{eV.{\AA}^{-1}}$ are set to obtain the accurate results.
A  vacuum of more than 16 $\mathrm{{\AA}}$ is introduced
to avoid the interaction between neighboring images.
\begin{figure*}
  \includegraphics[width=14cm]{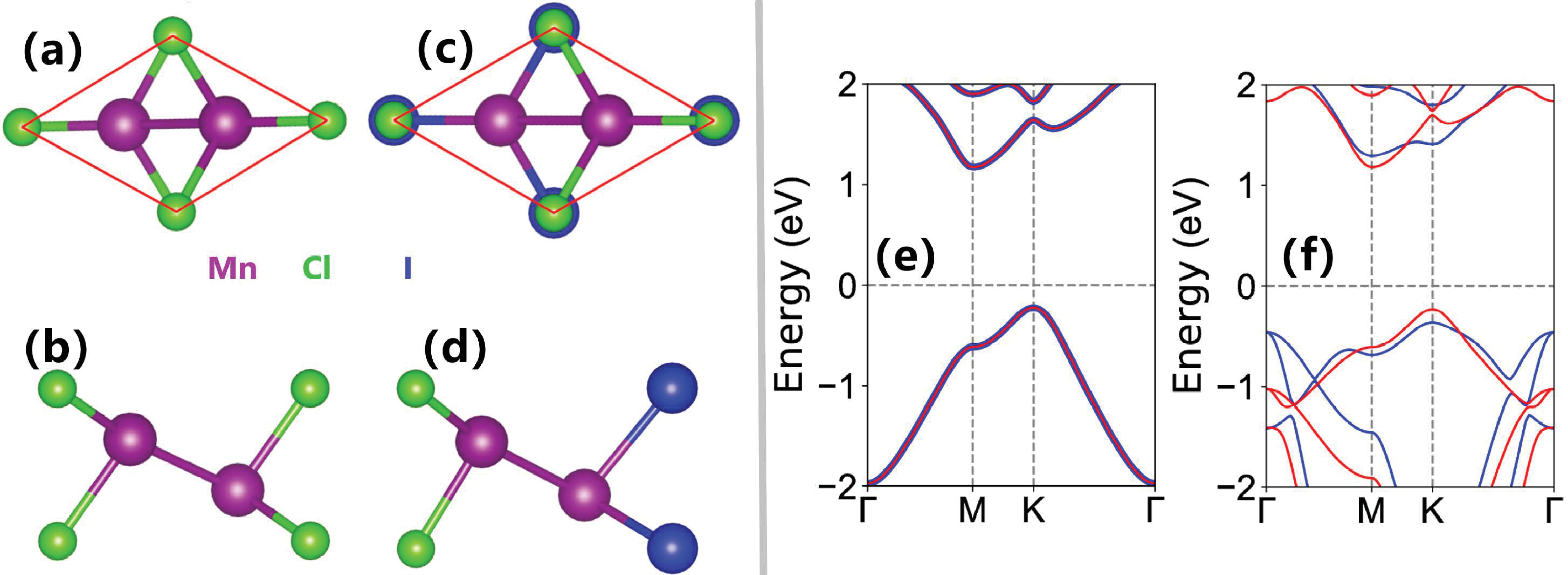}
  \caption{(Color online) The top (a,c) and side (b,d) views of the  crystal structures for  $\mathrm{Mn_2Cl_2}$ (a,b) and $\mathrm{Mn_2ClI}$ (c,d). The energy band structures of   $\mathrm{Mn_2Cl_2}$ (e) and $\mathrm{Mn_2ClI}$ (f), and the spin-up
and spin-down channels are depicted in blue and red.}\label{st2}
\end{figure*}
\begin{figure}
  \includegraphics[width=4cm]{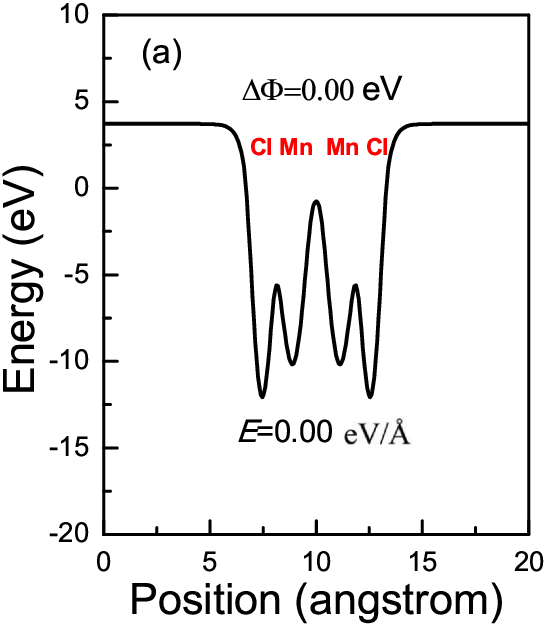}
  \includegraphics[width=4cm]{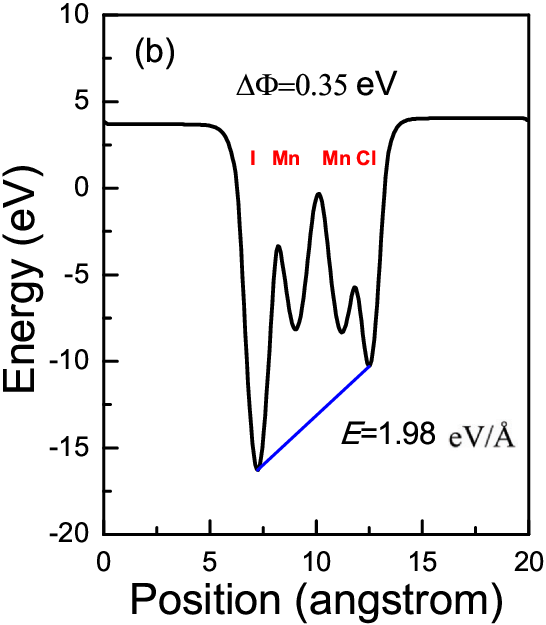}
  \caption{(Color online)For monolayer $\mathrm{Mn_2Cl_2}$ (a) and  $\mathrm{Mn_2ClI}$ (b), the planar averaged electrostatic potential energy variation along $z$ direction. $\Delta\Phi$ is the potential energy difference across the layer. $E$ means built-in electric field. Nonzero built-in electric field produces the spin splitting.}\label{vot}
\end{figure}

\section{1. $\mathrm{V_2SSeO}$ monolayer }
The $\mathrm{V_2Se_2O}$ monolayer has been predicted to be a 2D altermagnetic material\cite{k13}, which
possesses three atomic layers with
two V atoms and one O atom in the middle layer sandwiched by
two Se  layers (see \autoref{st1} (a) and (b)). Each V atom is surrounded by four Se atoms, and  the two V atoms  in the primitive cell
form  AFM  configuration. The two V atoms have the same surrounding Se atomic arrangement as a rectangle, but  the directions  of the two rectangle are different ($x$ and $y$ directions), which leads to the spin splitting.
However, two  V-atom related  sublattices  is related by a diagonal mirror symmetry $M_{xy}$ (see \autoref{st1-1}), and then two spin channels in the band structures is also related by $M_{xy}$.  It is noted that two V-atom related  sublattices cannot be transformed to each other by any
translation operation.

Janus monolayer $\mathrm{V_2SSeO}$   possesses similar crystal structures with $\mathrm{V_2Se_2O}$ (see \autoref{st1} (c) and (d)),  which can be constructed  by  replacing one of two Se layers with S atoms in monolayer  $\mathrm{V_2Se_2O}$.   Each V atom in $\mathrm{V_2SSeO}$ is surrounded by two S and two Se atoms. The two V atoms have the same surrounding  atomic arrangement of S and Se  as a rectangle with different orientation. The two  V-atom related  sublattices  is related by  $M_{xy}$. So, the energy band structures between $\mathrm{V_2SSeO}$ and $\mathrm{V_2Se_2O}$ should be similar.

A Hubbard correction $U_{eff}$=4.3 eV\cite{k13} is used to perform the related calculations, and the optimized lattice constants $a$=$b$=3.88 $\mathrm{{\AA}}$ for  $\mathrm{V_2SSeO}$. The AFM energy of $\mathrm{V_2SSeO}$ is 0.582 eV  lower than that of FM case.
The  magnetic moments of  two V atoms are 1.833 $\mu_B$ and -1.838 $\mu_B$, and total magnetic moment  is strictly 0.00 $\mu_B$.
The energy band structures of both $\mathrm{V_2Se_2O}$ and $\mathrm{V_2SSeO}$ are plotted in \autoref{st1} (e) and (f), and they show the similar energy band structures with obvious spin splitting. It is clearly seen that there are two equivalent valleys at high-symmetry  X and Y  points for both valence  and conduction  bands as valence band maximum (VBM) and conduction band bottom (CBM), which are  related by $M_{xy}$. States around X and Y points are mainly from two different V atoms with opposite spins, producing altermagnetism in the absence of SOC. The  spin-valley locking can be observed, which can be tuned by uniaxial strain\cite{k13}. For example, a compressive/tensile strain along $x$ direction
makes the energies of X and Y valleys be different, producing the spin-valley polarization.

\begin{figure*}
  \includegraphics[width=14cm]{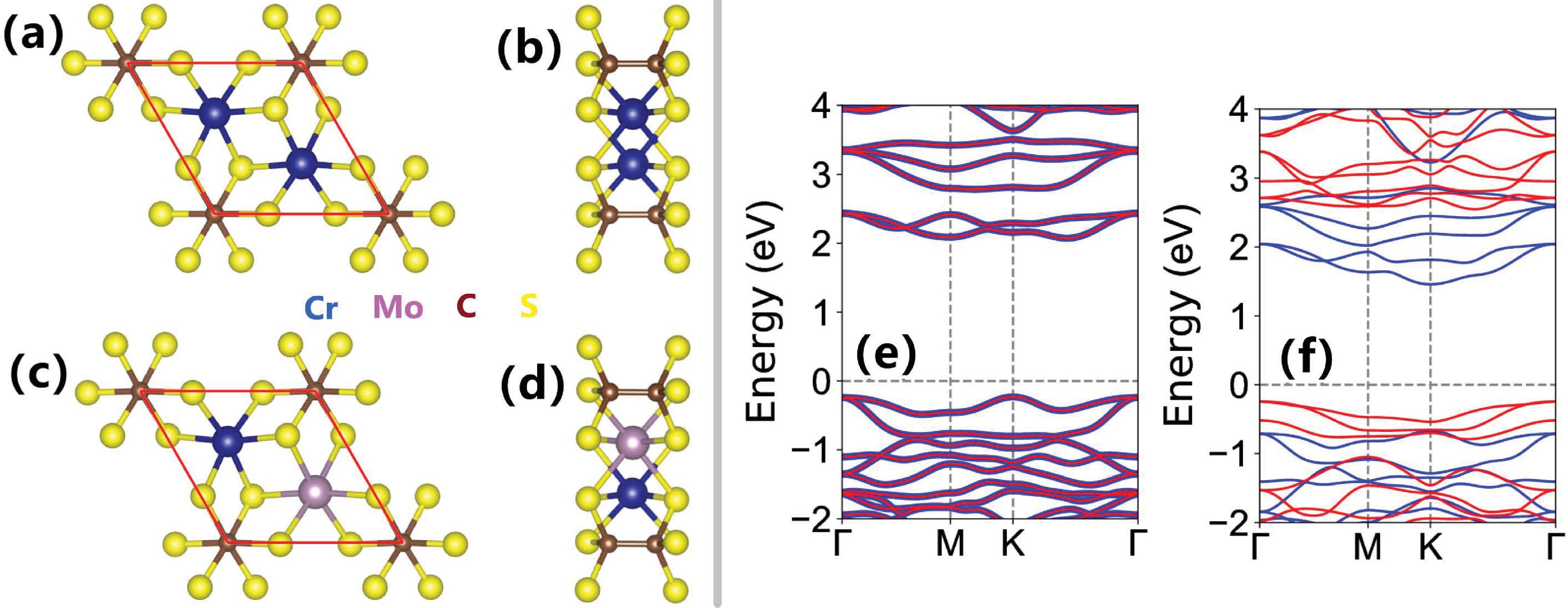}
  \caption{(Color online) The top (a,c) and side (b,d) views of the  crystal structures for  $\mathrm{Cr_2C_2S_6}$ (a,b) and $\mathrm{CrMoC_2S_6}$ (c,d). The energy band structures of    $\mathrm{Cr_2C_2S_6}$ (e) and $\mathrm{CrMoC_2S_6}$ (f), and the spin-up
and spin-down channels are depicted in blue and red.}\label{st3}
\end{figure*}
\begin{figure}
  \includegraphics[width=6cm]{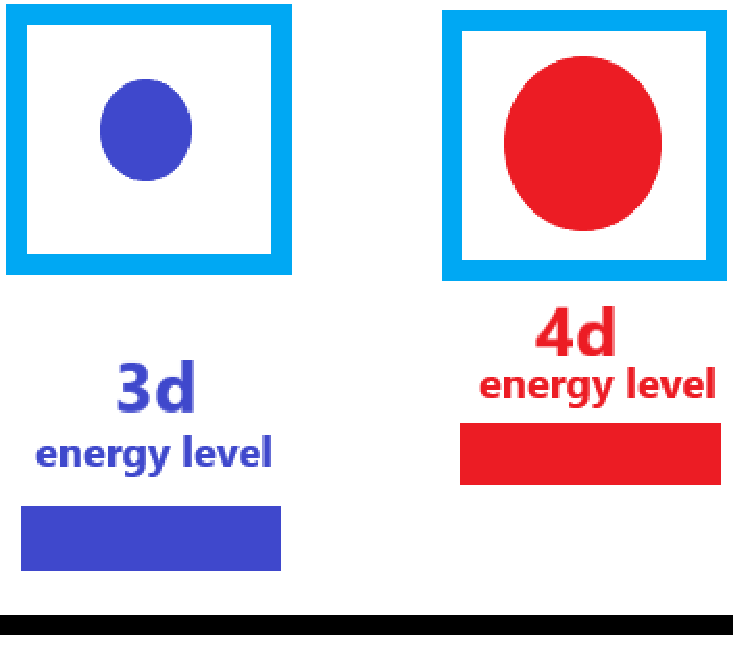}
  \caption{(Color online)The two magnetic atoms are 3$d$ Cr and 4$d$ Mo, respectively. They have the same  surrounding atomic arrangement, but the energy level of 4$d$ electron is higher than that of 3$d$ one, producing the spin splitting.}\label{st3-1}
\end{figure}

\begin{figure*}
  \includegraphics[width=14cm]{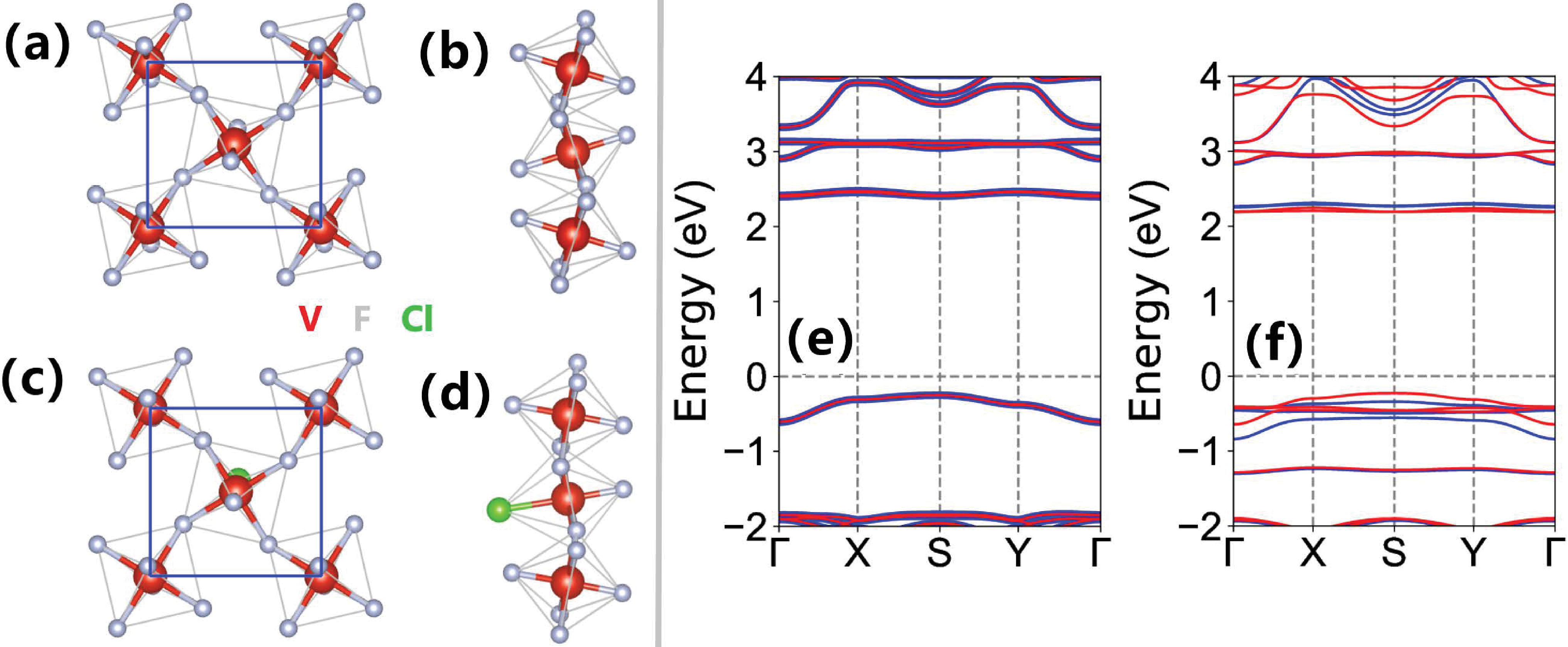}
  \caption{(Color online) The top (a,c) and side (b,d) views of the  crystal structures for  $\mathrm{V_2F_8}$ (a,b) and $\mathrm{V_2F_7Cl}$ (c,d). The energy band structures of     $\mathrm{V_2F_8}$ (e) and $\mathrm{V_2F_7Cl}$ (f), and the spin-up
and spin-down channels are depicted in blue and red.}\label{st4}
\end{figure*}
\begin{figure}
  \includegraphics[width=6cm]{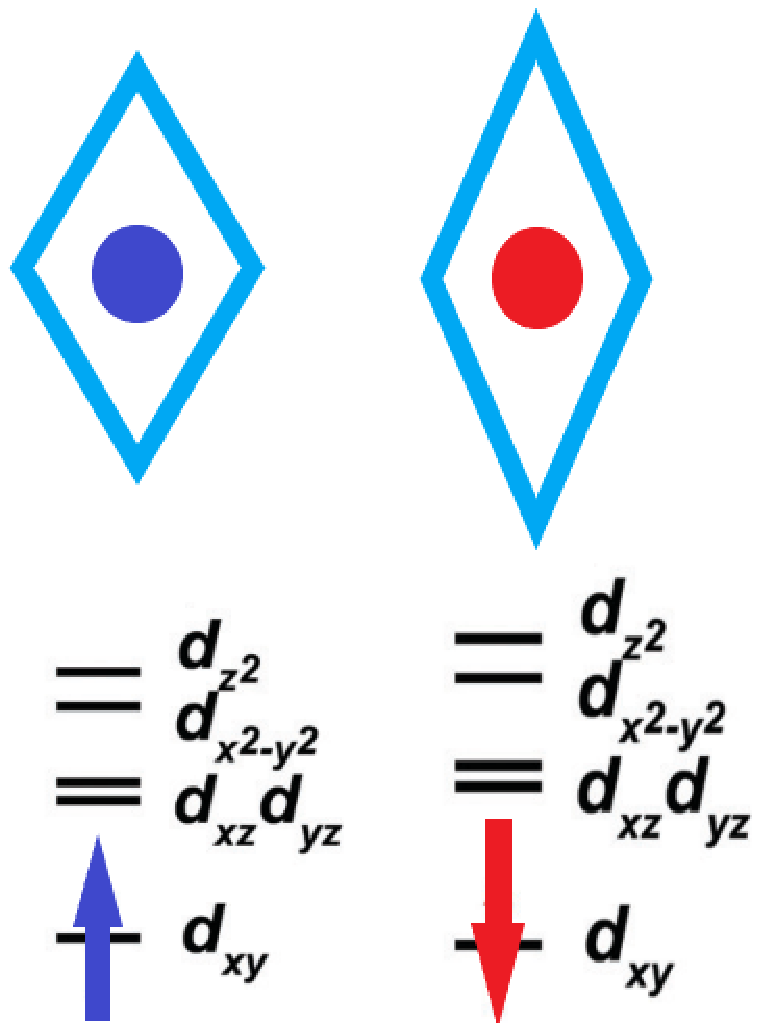}
  \caption{(Color online)The two V atoms have different levels of Jahn-Teller
distortion, which leads to that the $d$ orbitals of two V atoms have different splitting sizes,  producing the spin splitting in the band structures.}\label{st4-1}
\end{figure}
\section{2. $\mathrm{Mn_2ClI}$ monolayer}
The monolayer $\mathrm{Mn_2Cl_2}$ is predicted to be a A-type AFM semiconductors without spin splitting, and  the spin polarization  can be achieved by external electric field\cite{k15}. $\mathrm{Mn_2Cl_2}$  is composed of two
hexagonal Mn layers sandwiched by two hexagonal Cl
layers in the stacking order of Cl-Mn-Mn-Cl with both magnetic atomic layers bonded with metallic bonds
and ionic bonds (see \autoref{st2} (a) and (b)).  The two Mn atoms have the same surrounding  atomic arrangement of Cl and Mn, which gives rise to the degeneration of electron spin.

Janus monolayer $\mathrm{Mn_2ClI}$   also  consists  of four atomic layers in the sequence of Cl-Mn-Mn-I (see \autoref{st2} (c) and (d)), which can be constructed  by  replacing one of two Cl layers with I atoms in monolayer  $\mathrm{Mn_2Cl_2}$.  The two Mn atoms have the different surrounding  atomic arrangement.  One Mn atom is sandwiched by
 Cl and Mn  layers, while another Mn atom is  sandwiched by
 I and Mn  layers, which can produce spin splitting. Similar to  $\mathrm{Mn_2ClF}$\cite{k14},  an intrinsic polar electric field along the $z$ direction can be induced due to the  different electronegativity of the Cl and I elements, which can realize EPD-AFM.

 We use a Hubbard correction $U_{eff}$=4.0 eV  to perform the related calculations, and the optimized lattice constants $a$=$b$=3.83 $\mathrm{{\AA}}$ for  $\mathrm{Mn_2ClI}$. The AFM ordering is the ground state, and its energy  is 0.535 eV  lower than  FM energy.
The  magnetic moments of two Mn atoms are 4.57 $\mu_B$ and -4.62 $\mu_B$ with  total magnetic moment per unit cell being strictly 0.00 $\mu_B$.
The energy band structures of both $\mathrm{Mn_2Cl_2}$ and $\mathrm{Mn_2ClI}$ are shown in \autoref{st2} (e) and (f). It is clearly seen that no spin splitting can be observed for $\mathrm{Mn_2Cl_2}$, while $\mathrm{Mn_2ClI}$ shows obvious spin splitting.
This difference is because the $\mathrm{Mn_2ClI}$ possesses the out-of-plane polar electric field, while the built-in electric field of $\mathrm{Mn_2Cl_2}$ disappears.  To clearly identify the inherent electric field of $\mathrm{Mn_2ClI}$, the planar  average of the
electrostatic potential  energy  along $z$ direction is plotted in \autoref{vot} along with that of $\mathrm{Mn_2Cl_2}$. The mirror asymmetry can induce  an electrostatic potential gradient ($\Delta\Phi$) of about 0.35 eV for $\mathrm{Mn_2ClI}$, which is related to the work function change of the structure.
The magnitude of the net vertical electric field can be estimated by the
slope of the plane-averaged electrostatic potential between top  and bottom atoms' minima. For $\mathrm{Mn_2Cl_2}$, the built-in electric field is 0.00 $\mathrm{eV/{\AA}}$ due to out-of-plane structural symmetry, and is 1.98  $\mathrm{eV/{\AA}}$ for $\mathrm{Mn_2ClI}$ with structural asymmetry.
So, the $\mathrm{Mn_2ClI}$ possesses EPD-AFM.
   Calculated results show that $\mathrm{Mn_2ClI}$ is  an indirect band gap semiconductor (1.417 eV)  with  VBM and CBM  at high symmetry K and  M points, respectively.  The VBM and CBM are provided by the same spin-up channel. These results of $\mathrm{Mn_2ClI}$ are similar to those of   $\mathrm{Mn_2ClF}$\cite{k14}.

\section{3. $\mathrm{CrMoC_2S_6}$ monolayer}
Inspired by the successful synthesis of the 2D  $\mathrm{Mn_2P_2S_6}$
monolayer with AFM ordering\cite{k17}, we construct the $\mathrm{Cr_2C_2S_6}$, whose crystal structures are shown in \autoref{st3} (a) and (b). There are two Cr, two C and
six S  atoms in the primitive cell.
The magnetic
Cr atoms form a honeycomb lattice within the same layer.
The two Cr atoms in the primitive cell are  coordinated by a  octahedron of six
S  atoms from the neighboring ligands 3S-2C-3S
with the centers of the hexagons occupied by the C-C units.
The two Cr atoms have the same octahedral environment, producing the degeneration of electron spin.
To break the equivalence of two magnetic atoms, 2D $\mathrm{CrMoC_2S_6}$
can be built by substituting one Cr in the $\mathrm{Cr_2C_2S_6}$
via isovalent Mo, which are plotted in \autoref{st3} (c) and (d).
The two magnetic atoms are 3$d$ Cr and 4$d$ Mo, respectively,  which  have the same  surrounding atomic arrangement (see \autoref{st3-1}). However, the energy level of 4$d$ electron is higher than that of 3$d$ one, which will lead to spin splitting.

 The GGA+$U$ method with a Hubbard $U_{eff}$=3 eV for both Cr and Mo\cite{k18} is employed for
the related calculations. The optimized lattice constants $a$=$b$=5.714 $\mathrm{{\AA}}$, which agrees well with previous theoretical value\cite{k18}. The AFM ordering is the ground state with energy difference being -0.580 eV between AFM and  FM cases.
The  magnetic moments of Cr and Mo atoms are 2.88  $\mu_B$ and -2.34 $\mu_B$, respectively. However, the  total magnetic moment per unit cell is strictly 0.00 $\mu_B$.
The energy band structures of both $\mathrm{Cr_2C_2S_6}$ and $\mathrm{CrMoC_2S_6}$ are shown in \autoref{st3} (e) and (f). Calculated results show  no spin splitting for $\mathrm{Cr_2C_2S_6}$, and  obvious spin splitting for $\mathrm{CrMoC_2S_6}$. The spin-down channel from 4$d$ orbital character move toward higher energy than spin-up channel from 3$d$ one.
 The $\mathrm{CrMoC_2S_6}$ is a bipolar antiferromagnetic semiconductor with VBM and
CBM in different spin channels, and the gap value is 1.702 eV.
Compared with  $\mathrm{V_2SSeO}$ and $\mathrm{Mn_2ClI}$, $\mathrm{CrMoC_2S_6}$ possesses larger spin splitting and more obvious spontaneous spin polarization.
In general, 4$d$ electrons have weaker electron correlation than 3$d$ electrons.  A Hubbard correction $U_{eff}$=3.0/2.0 eV for Cr/Mo is adopted   to recalculate the related results. The magnetic ground state is still AFM ordering, and the energy band structures still show obvious spin splitting.

\section{4. $\mathrm{V_2F_7Cl}$ monolayer}
Monolayer  $\mathrm{V_2F_8}$ is predicted to have
antiferromagnetic ferroelasticity and bidirectional
negative Poisson's ratio\cite{k19}. The two V atoms  are  coordinated by a  octahedron of six
F  atoms (see \autoref{st4} (a) and (b)).  According to the following calculation results,  the V-F
bonds (1.947 $\mathrm{{\AA}}$) within the $xy$-plane are significantly longer than those (1.747 $\mathrm{{\AA}}$)
along the $z$-direction, producing Jahn-Teller
distortion, which splits the $d$ orbitals into
four subgroups: the $d_{xy}$, $d_{x^2-y^2}$, and $d_{z^2}$ singlets, and the
$d_{xz}$ and $d_{yz}$ doublet, with the lowest $d_{xy}$ orbital occupied\cite{k19}.
For net in-plane environment, the two V atoms are equivalent, leading to  the degeneration of electron spin.
To break the equivalence of two magnetic atoms, 2D $\mathrm{V_2F_7Cl}$
is constructed  by substituting one F  in the $\mathrm{V_2F_8}$ with Cl (see \autoref{st4} (c) and (d)).
The two V atoms in $\mathrm{V_2F_7Cl}$  have different levels of Jahn-Teller
distortion, which leads to that the $d$ orbitals of two V atoms have different splitting sizes (see \autoref{st4-1}),  producing the spin splitting in the band structures.

The GGA+$U$ method with a Hubbard $U_{eff}$=3 eV for V atom  is adopted for
the related calculations. The optimized lattice constants $a$=$b$=5.366 $\mathrm{{\AA}}$ for $\mathrm{V_2F_7Cl}$. The AFM ordering is the ground state, and  the  energy difference is -0.014 eV between AFM and   FM orderings.
The  magnetic moments of two V  atoms are 1.00  $\mu_B$ and -1.05 $\mu_B$, and the  total magnetic moment per unit cell is strictly 0.00 $\mu_B$.
The energy band structures of both $\mathrm{V_2F_8}$ and $\mathrm{V_2F_7Cl}$  are plotted in \autoref{st4} (e) and (f). Calculated results show  no spin splitting for $\mathrm{V_2F_8}$, and  observable spin splitting for $\mathrm{V_2F_7Cl}$.
 The $\mathrm{V_2F_7Cl}$ is an indirect semiconductor (2.416 eV) with VBM (S point) and
CBM ($\Gamma$ point) in the same spin channels .

\section{Discussion and Conclusion}
In fact, the four types can coexist with each other in a material. For example a 2D altermagnet, if the magnetic atoms have opposite layer spin polarization (A-type AFM ordering) with an out-of-plane built-in electric field,  the spontaneous spin-valley polarization can be produced\cite{k14}.
The magnetic atoms with opposite layer spin polarization   have been predicted  in 2D $\mathrm{Ca(CoN)_2}$ without built-in electric field\cite{yz}.
So, it is possible to find electric-potential-difference altermagnet (EPD-AM).

 In summary,  we propose an intuitional strategy to find or produce  spin-splitting in AFM   materials without SOC.
It is demonstrated that four different types of 2D AFM materials ($\mathrm{V_2SSeO}$,  $\mathrm{Mn_2ClI}$,  $\mathrm{CrMoC_2S_6}$ and  $\mathrm{V_2F_7Cl}$) can be  used to confirm our proposal.
For $\mathrm{V_2SSeO}$, the different environment can be induced by the orientation of surrounding atoms arrange.
For  $\mathrm{Mn_2ClI}$, the out-of-plane  inherent electric field leads to different environment.
For $\mathrm{CrMoC_2S_6}$, 4$d$ electron has higher energy than 3$d$ electron, which produces different environment.
For $\mathrm{V_2F_7Cl}$, different levels of Jahn-Teller
distortion gives rise to different environment. To produce spin splitting in AFM materials, other different mechanisms can  be found, but the essential reason is that the magnetic atoms with opposite spin polarization locate in the different environment (surrounding atomic arrangement).

\begin{acknowledgments}
Y.S.A. is supported by the Singapore Ministry of Education Academic Research Fund Tier 2 (Award No. MOE-T2EP50221-0019). We are grateful to Shanxi Supercomputing Center of China, and the calculations were performed on TianHe-2.
\end{acknowledgments}

\end{document}